\documentclass[structabstract]{aa}  
\usepackage{amsmath}
\usepackage{graphicx}
\usepackage{txfonts}
\usepackage{natbib,twoopt}
\usepackage{color}
\usepackage{multirow}

\bibpunct{(}{)}{;}{a}{}{,} 
\newcommandtwoopt{\citeads}[3][][]{\href{http://adsabs.harvard.edu/abs/#3}%
{\citealp[#1][#2]{#3}}} 
\newcommandtwoopt{\citepads}[3][][]{\href{http://adsabs.harvard.edu/abs/#3}%
{\citep[#1][#2]{#3}}} 
\newcommandtwoopt{\citetads}[3][][]{\href{http://adsabs.harvard.edu/abs/#3}%
{\citet[#1][#2]{#3}}}
\newcommandtwoopt{\citeyearads}[3][][]%
{\href{http://adsabs.harvard.edu/abs/#3}{\citeyear[#1][#2]{#3}}}

\begin{document}

	\title{The Araucaria Project: High-precision orbital parallax and masses of the eclipsing binary TZ~Fornacis\thanks{Based on observations made with ESO telescopes at Paranal observatory under program IDs 094.D-0320}}
	\titlerunning{High-precision orbital parallax and masses of the eclipsing binary TZ~Fornacis}

	\author{ A.~Gallenne\inst{1,2},
					G.~Pietrzy\'nski\inst{1,3},
					D.~Graczyk\inst{1,4},
					P.~Konorski\inst{3},
					P.~Kervella\inst{5, 6},
					A.~M\'erand\inst{2},
				W.~Gieren\inst{1,4},
				 R.~I.~Anderson\inst{7,8}
				\and S.~Villanova\inst{1}
  				}
  				  				
  	\authorrunning{A. Gallenne et al.}

\institute{Universidad de Concepci\'on, Departamento de Astronom\'ia, Casilla 160-C, Concepci\'on, Chile
		\and European Southern Observatory, Alonso de C\'ordova 3107, Casilla 19001, Santiago 19, Chile
  	\and Warsaw University Observatory, Al. Ujazdowskie 4, 00-478, Warsaw, Poland
  	\and Millenium Institute of Astrophysics, Av. Vicu{\~n}a Mackenna 4860, Santiago, Chile
    \and LESIA, Observatoire de Paris, CNRS UMR 8109, UPMC, Universit\'e Paris Diderot, 5 Place Jules Janssen, F-92195 Meudon, France
    \and Unidad Mixta Internacional Franco-Chilena de Astronom\'ia, CNRS/INSU, France (UMI 3386) and Departamento de Astronom\'ia, Universidad de Chile, Camino El Observatorio 1515, Las Condes, Santiago, Chile
	\and Department of Physics and Astronomy, Johns Hopkins University, Baltimore, MD 21218, USA
	\and D\'epartement d'Astronomie, Universit\'e de Gen\`eve, 51 Ch. des Maillettes, 1290 Sauverny, Switzerland
  	}
  
  \offprints{A. Gallenne} \mail{agallenne@astro-udec.cl}

   \date{Received June 17, 2015; accepted November 13, 2015}

 
  \abstract
   {Independent distance estimates are particularly useful to check the precision of other distance indicators, while accurate and precise masses are necessary to constrain evolution models.}
   {The goal is to measure the masses and distance of the detached eclipsing-binary TZ~For with a precision level lower than 1\,\% using a fully  geometrical and empirical  method.}
   {We obtained the first interferometric observations of TZ~For with the VLTI/PIONIER combiner, which we combined with new and precise radial velocity measurements to derive its three-dimensional orbit, masses, and distance.}
   {The system is well resolved by PIONIER at each observing epoch, which allowed a combined fit with eleven astrometric positions. Our derived values are in a good agreement with previous work, but with an improved precision. We measured the mass of both components to be $M_1 = 2.057 \pm 0.001\,M_\odot$ and $M_2 = 1.958 \pm 0.001\,M_\odot$. The comparison with stellar evolution models gives an age of the system of $1.20 \pm 0.10$\,Gyr. We also derived the distance to the system with a precision level of 1.1\,\%:  $d = 185.9 \pm 1.9$\,pc. Such precise and accurate geometrical distances to eclipsing binaries provide a unique opportunity to test the absolute calibration of the surface brightness-colour relation for late-type stars, and will also provide the best opportunity to check on the future Gaia measurements for possible systematic errors.}
   {}

 
 \maketitle

%

\section{Introduction}

In the course of the Araucaria project, we applied different techniques for distance measurement in order to track down the influence of the population effects on the most important standard candles like Cepheids, RR Lyrae stars, red clump stars, tip of the red-giant branch (TRGB), etc. \citep{Gieren_2005_09_0,Gieren_2005_08_0}. Binary systems are of particular importance in our project.

Binary stars are the only tool enabling direct and accurate stellar mass measurements. This parameter is fundamental in order to understand the structure and evolution of stars, and to check the consistency with predicted values from theoretical models. In addition, such systems also provide geometrical distance measurements, which are particularly important for the cosmic distance scale. Eclipsing binary systems have the potential to provide the most accurate distance determinations to nearby galaxies with an unprecedented precision of about 1\,\%. \citet{Pietrzynski_2013_03_0} and \citet{Graczyk_2014_01_0} have already proved their efficiency by measuring the distance to the Large and Small Magellanic Cloud (LMC, SMC) with a precision of 2\,\% and 3\,\%, respectively. The approach is to fit high-quality spectroscopic and photometric observations to obtain very accurate masses, sizes, and surface brightness ratios of the components, and using a surface brightness-colour relation (SBCR, with the $V-K$ colour), the distance to the system. The results of \citet{Pietrzynski_2013_03_0}, who used eight eclipsing binary systems in the LMC, currently provide the best zero point for the whole extragalactic distance scale. However, the total error budget is dominated by the precision of the SBCR (about 2\%). Therefore, a significant improvement on the distance determination with this technique would only be possible with a better calibration of this relation.

Another approach for measuring stellar parameters and geometrical distances with binary systems at 1 \% accuracy is to combine spectroscopic and astrometric observations \citep[see e.g.][]{Torres_2009_08_0, Zwahlen_2004_10_0}. This method provides a direct geometric distance and model-independent masses, and does not require any assumptions. However, the systems need to be spatially resolved to enable astrometric measurements, which is not always the case with single-dish telescope observations where the components are too close. A higher angular resolution is provided from long-baseline interferometry (LBI), where close binary systems ($< 20$\,mas) can be resolved. LBI already proved its efficiency in terms of angular resolution and accuracy for close binary stars \citep[see e.g.][]{Gallenne_2015_07_0,Gallenne_2014_01_0,Gallenne_2013_04_0,Le-Bouquin_2013_03_0,Baron_2012_06_0}. 

The nearby eclipsing binary system \object{TZ~Fornacis} (HD~20301) is an evolved double-line spectroscopic system composed of a giant (a G8III primary, $V = 7.51$\,mag) and a subgiant star (a F7III secondary, $V = 7.76$\,mag) with a 75.7-days circular orbit. Absolute parameters of this system were derived by \citet{Andersen_1991_06_0} from the analysis of spectroscopic and photometric observations with an average accuracy of 2\,\% on the stellar parameters. Long-baseline interferometry can provide better accuracy with high-precision measurements, and is the only observing technique able to spatially resolve this system as the angular semi-major axis is $\sim 3.0$\,mas. This system  provides us a unique opportunity to apply both  approaches to the same system, and to perform several interesting tests. In particular, comparison of the distance obtained based on spectroscopic and interferometric observations to the corresponding  distance derived through the analysis of the radial velocity and photometric light curves together with a SBCR will allow the absolute calibration of the SBCR for late-type stars to be tested. Since the SBCR is the heart of accurate distance determination with eclipsing binaries to nearby galaxies, this test performed at the level of a very few percentage points is of great importance for the extragalactic distance scale. 

Recently, we obtained new high-quality interferometric and spectroscopic data for  this system in the course of the Araucaria project. The detailed comparison of the results from both techniques will be presented in the future once we collect the near infrared photometry. In this paper we present results from the analysis of the interferometric and spectroscopic data and compare them with the results of \citet{Andersen_1991_06_0}.

The spectroscopic and interferometric observations and data reduction process are presented in Sect.~\ref{section__observations_and_data_reduction}. In Sect.~\ref{section__orbital_solutions_masses_and_distance}, we performed a combined fit to derive all the orbital elements, the masses, and the orbital parallax. We then discuss our results in Sect.~\ref{section__discussion} before concluding in Sect.~\ref{section__conclusion}.

\section{Observations and data reduction}
\label{section__observations_and_data_reduction}

\begin{table}[!h]
\centering
\caption{Measured radial velocities of both components.}
\begin{tabular}{cccccc}
	\hline\hline
	   MJD    & $V_1$  & $e_\mathrm{V1}$ & $V_2$  & $e_\mathrm{V2}$ & Instrument \\
	          & (km/s) &     (km/s)      & (km/s) &     (km/s)      &  \\ \hline
54741.186  &  53.83  &  0.03  &  -19.20  &  0.11  & CORALIE  \\
54742.184  &  52.47  &  0.03  &  -17.93  &  0.11  & CORALIE  \\
54743.161  &  50.89  &  0.03  &  -16.25  &  0.11  & CORALIE  \\
54743.187  &  50.85  &  0.03  &  -16.04  &  0.11  & CORALIE  \\
54744.092  &  49.20  &  0.03  &  -14.27  &  0.11  & CORALIE  \\
54744.169  &  49.03  &  0.03  &  -14.15  &  0.11  & CORALIE  \\
54744.384  &  48.62  &  0.03  &  -13.65  &  0.11  & CORALIE  \\
54745.111  &  47.11  &  0.03  &  -12.22  &  0.11  & CORALIE  \\
54746.115  &  44.88  &  0.03  &  -9.96  &  0.11  & CORALIE  \\
54746.233  &  44.61  &  0.03  &  -9.61  &  0.11  & CORALIE  \\
54747.182  &  42.28  &  0.03  &  -7.25  &  0.11  & CORALIE  \\
54747.290  &  42.01  &  0.03  &  -7.04  &  0.11  & CORALIE  \\
54747.392  &  41.73  &  0.03  &  -6.83  &  0.11  & CORALIE  \\
54819.102  &  50.38  &  0.03  &  -15.74  &  0.11  & CORALIE  \\
54820.110  &  48.48  &  0.03  &  -13.77  &  0.11  & CORALIE  \\
54821.093  &  46.41  &  0.03  &  -11.64  &  0.11  & CORALIE  \\
54822.117  &  44.05  &  0.03  &  -9.10  &  0.11  & CORALIE  \\
54887.047  &  56.78  &  0.03  &  -22.56  &  0.11  & HARPS  \\
55060.258  &  10.61  &  0.03  &  --  &  --  & HARPS  \\
55086.245  &  -9.71  &  0.03  &  47.35  &  0.11  & CORALIE  \\
55086.321  &  -9.61  &  0.03  &  47.37  &  0.11  & CORALIE  \\
55087.190  &  -7.57  &  0.03  &  45.09  &  0.11  & CORALIE  \\
55087.269  &  -7.33  &  0.03  &  45.09  &  0.11  & CORALIE  \\
55088.204  &  -4.99  &  0.03  &  42.51  &  0.11  & CORALIE  \\
55088.291  &  -4.74  &  0.03  &  42.20  &  0.11  & CORALIE  \\
55089.200  &  -2.29  &  0.03  &  --  &  --  & CORALIE  \\
55089.278  &  -2.06  &  0.03  &  39.30  &  0.11  & CORALIE  \\
55090.187  &  0.52  &  0.03  &  --  &  --  & CORALIE  \\
55090.258  &  0.70  &  0.03  &  --  &  --  & CORALIE  \\
55144.256  &  -12.15  &  0.03  &  49.89  &  0.11  & HARPS  \\
55431.310  &  33.62  &  0.03  &  --  &  --  & HARPS  \\
55479.107  &  33.59  &  0.03  &  --  &  --  & HARPS  \\
55499.146  &  52.00  &  0.03  &  -17.33  &  0.11  & CORALIE  \\
55500.097  &  50.39  &  0.03  &  -15.57  &  0.11  & CORALIE  \\
55501.349  &  48.02  &  0.03  &  -13.12  &  0.11  & CORALIE  \\
55504.139  &  41.52  &  0.03  &  -6.35  &  0.11  & HARPS  \\
55534.026  &  -19.58  &  0.03  &  57.76  &  0.11  & HARPS  \\
55536.020  &  -17.40  &  0.03  &  55.50  &  0.11  & HARPS  \\
56213.191  &  -20.72  &  0.03  &  58.96  &  0.11  & HARPS  \\
56213.351  &  -20.65  &  0.03  &  58.94  &  0.11  & HARPS  \\
56214.377  &  -20.10  &  0.03  &  58.33  &  0.11  & HARPS  \\
56242.019  &  49.20  &  0.03  &  -14.34  &  0.11  & HARPS  \\
56552.349  &  56.94  &  0.03  &  -22.64  &  0.11  & HARPS  \\
56577.407  &  -0.83  &  0.03  &  38.36  &  0.11  & HARPS  \\
56578.409  &  -3.64  &  0.03  &  41.26  &  0.11  & HARPS  \\
56826.423  &  -10.05  &  0.03  &  47.77  &  0.11  & CORALIE  \\
56827.420  &  -7.71  &  0.03  &  45.53  &  0.11  & CORALIE  \\
56876.440  &  10.05  &  0.03  &  --  &  --  & HARPS  \\
56877.433  &  6.94  &  0.03  &  --  &  --  & HARPS  \\
56878.440  &  3.87  &  0.03  &  --  &  --  & HARPS  \\
56879.436  &  0.92  &  0.03  &  --  &  --  & HARPS  \\
56908.315  &  6.96  &  0.03  &  --  &  --  & HARPS  \\
57005.023  &  56.70  &  0.03  &  -22.29  &  0.11  & HARPS  \\
57029.029  &  6.12  &  0.03  &  --  &  --  & HARPS  \\
\hline
\end{tabular}
\label{table__rv}
\end{table}

\subsection{New spectroscopic observations}

We collected high-resolution echelle spectra from the instruments Euler/CORALIE and 3.6\,m/HARPS located at La Silla Observatory (Chile). CORALIE \citep{Queloz_2001_09_0,Segransan_2010_02_0} offers a spectral resolution $R \sim 60~000$, while we used the EGGS mode of HARPS \citep{Mayor_2003_12_0}, which provided a resolution of about 80~000. Both instruments cover the spectral range $3900-6900\,\AA$. Calibrated CORALIE spectra were obtained using the Geneva pipeline, and HARPS data were processed with the standard ESO/HARPS pipeline reduction package. The average signal-to-noise ratio obtained ranges from 47 to 136 (median = 96).

\begin{figure*}[!ht]
\centering
\resizebox{\hsize}{!}{\includegraphics{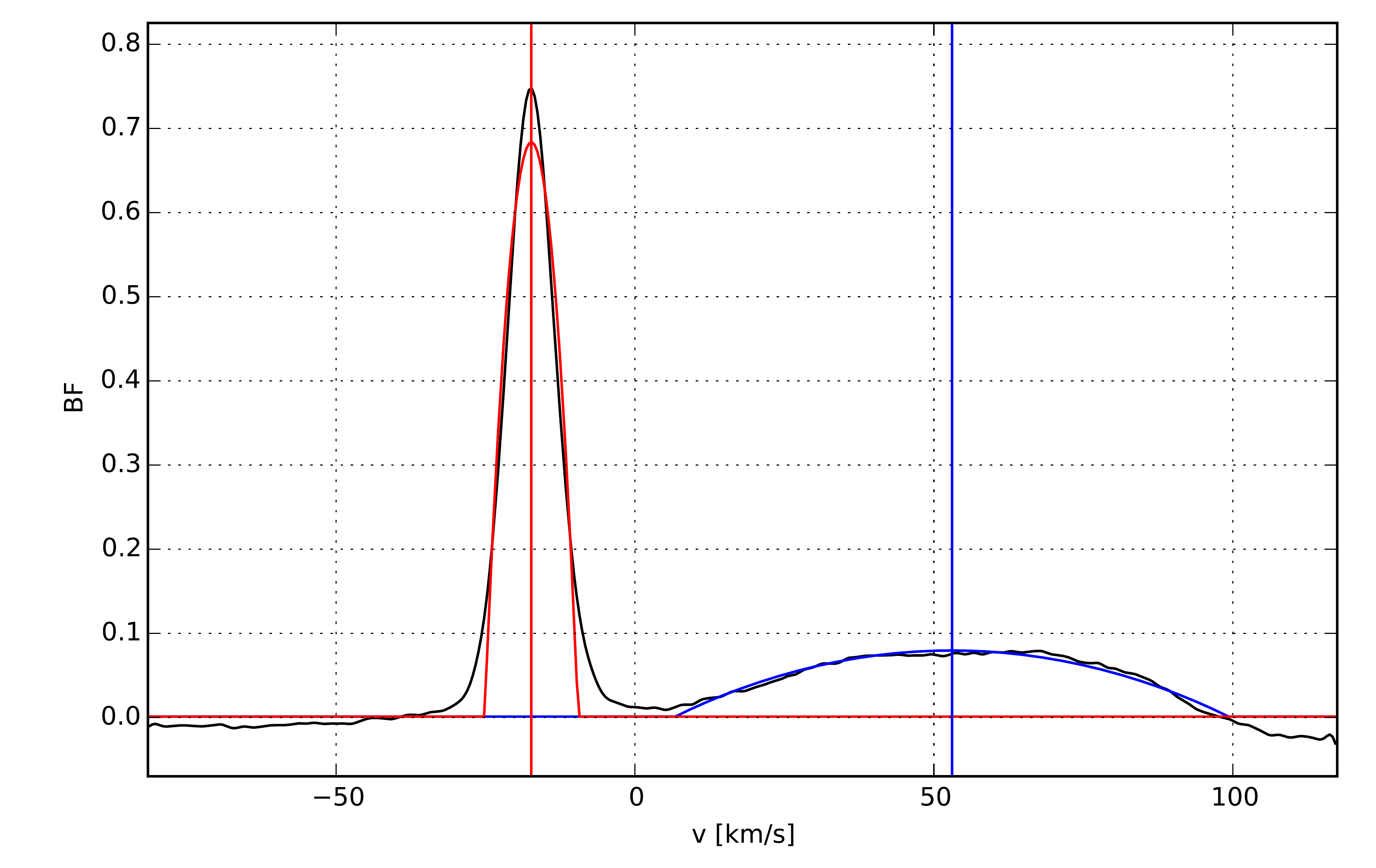}\includegraphics{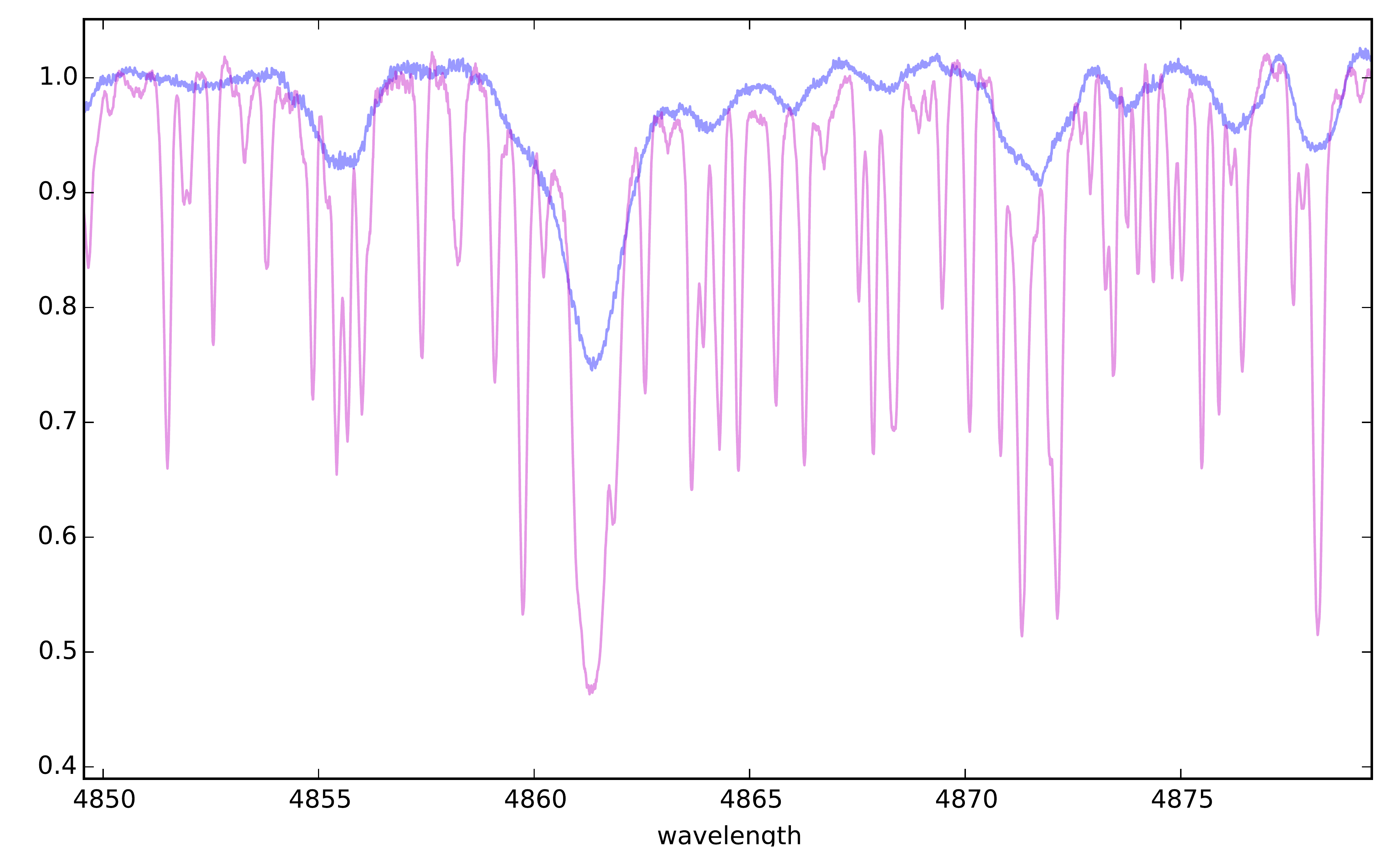}
}
\caption{{\it Left}: The broadening function profiles of TZ For derived from HARPS spectrum taken on December 6, 2010. The primary rotates synchronously (high peak on left) and the secondary's profile is flattened because of fast rotation. {\it Right}: normalized, disentangled spectra of the primary (narrow lines) and the secondary (broad shallow lines) around of H$\beta$ line. }
\label{figure_rv}
\end{figure*}

Radial velocities of both components were derived using the broadening function (BF) formalism \citep{Rucinski_1999__0,Rucinski_1992_11_0} implemented in the RaveSpan software \citep{Pilecki_2012_04_0}. As templates we used synthetic spectra from \citet{Coelho_2005_11_0} corresponding to atmospheric parameters of the TZ For components \citep{Andersen_1991_06_0}. Wavelength ranges used for RV determination were: 4125--4230, 4245--4320, 4350--4840, 4880--5290, 5350--5850, 5920--6250, 6300--6420 and 6600--6840\,\AA. In this way we excluded troublesome hydrogen lines and most of the strong telluric lines within the 4125--6840\,\AA\ range. We derived rotational broadenings $v_1\sin{i}=6.1 \pm 0.3\,$km~s$^{-1}$ and $v_2\sin{i}=45.7 \pm 1.0\,$km~s$^{-1}$ (see Fig.~\ref{figure_rv}). The BF profile of the secondary star (less massive, smaller, and hotter component) is highly rotationally broadened and the BF profiles of components are separated only in orbital quadratures.

To measure radial velocities we used two templates corresponding closely to temperature and gravity of the components. To account for line profiles blending we employed an iterative procedure. First we used the cooler template to determine velocities. We chose ten spectra uniformly distributed in orbital phase and measured the radial velocities of both components. Then we used a spectral decomposition method outlined by \citet{Gonzalez_2006_03_0} to separate the spectra. A decomposed spectrum of the primary was subsequently removed from all our spectra and we determined radial velocities of the secondary, but this time we used the hotter template, fixing radial velocities of the primary. Subsequently, we repeated the spectral decomposition using all of our spectra, removing the spectrum of one component when measuring the radial velocity of the other one, each time using the appropriate template (the cooler template for the primary and the hotter for the secondary). This way we overcame the formal limitation of the BF, which allows for the use of only one radial velocity template at a time. We repeated this process three times until no progress in minimizing  the dispersion of radial velocity curves was obtained. The disentangled spectra of the two stars are presented in Fig.~\ref{figure_rv}. The wavelength range of the disentangled spectra is 4125--6840\,\AA.

The resulting velocities are presented in Table~\ref{table__rv}. In the following, we used only these radial velocities for the fitting procedure as they were derived using the same method. This will allow an independent estimate of the parameters and check for possible systematics in our results.

The orbital period $P$ and the epoch of the primary spectroscopic conjunction $T_0$ were estimated from the radial velocity analysis and were used for the combined fit (Sect.~\ref{section__orbital_solutions_masses_and_distance}). We also detected a difference of 0.365\,km~s$^{-1}$ in the systemic velocity estimated from the two stars. After taking into account gravitational redshift on the surface of the components this difference is reduced to 0.208\,km~s$^{-1}$, but it is still significant. Although a complete study of this phenomena is beyond the scope of this paper, we note that a possible explanation can be the convective blueshift caused by cells in the atmosphere of the stars. In the case of TZ~For, the giant component seems be more affected by the blueshift than its companion. This shift is taken into account in the next section by fitting two different systemic velocities.

We also checked for systematic shifts in the determination of the semi-amplitude variables $K_1$ and $K_2$. Neglecting rotational and tidal distortion in the estimates of the radial velocities produces shifts of 1.4\,m~s$^{-1}$ and 5.5\,m~s$^{-1}$, respectively for $K_1$ and $K_2$. The light time effect contribution to the semi-amplitude are 2.5\,m~s$^{-1}$ for both components \citep[see][]{Konacki_2010_08_0}. These values were added to the respective velocity errors for the combined fit. Amplitude of periodic relativistic effects is below 1\,m~s$^{-1}$ so their contribution to the radial velocities error is negligible.

From the decomposed spectra, we derived the atmospheric parameters for both stars, although less precise for the secondary because of broader lines. They were obtained using equivalent widths (EQW) of FeI line. We measured EQWs using Gaussian fitting. The strong lines showing deviation from Gaussian shape were rejected. Initial temperatures of the atmospherical models were assumed to be those typical of the spectral type of the two stars. We then refined them during the abundance analysis. As a first step, atmospherical models were calculated using ATLAS9 models \citep{Kurucz_1970__0} using the initial estimates of $T_\mathrm{eff}$ and $v_\mathrm{t} = 1.0\,\mathrm{km~s^{-1}}$ (turbulent velocity), and [Fe/H] = 0. The value of $T_\mathrm{eff}$ and $v_\mathrm{t}$ were then adjusted and new atmospheric models calculated in an iterative way in order to remove trends in excitation potential (EP) and equivalent width vs. abundance for $T_\mathrm{eff}$ and $v_\mathrm{t}$, respectively. The effective gravities were fixed to the values from \citet{Andersen_1991_06_0} because they are better constrained. The [Fe/H] value of the models were changed at each iteration according to the output of the abundance analysis. The local thermodynamic equilibrium (LTE) program MOOG \citep{Sneden_1973_09_0} was used for the abundance analysis. During the analysis we also rejected the lines deviating more than 3$\sigma$ from the mean [Fe/H] abundance. The derived stellar parameters are listed in Table~\ref{table__Teff}. All parameters are consistent with previous work, but the effective temperature of the secondary do not have improved precision. A detailed abundance analysis will be published in a forthcoming paper about the detailed modeling of the stellar interior to calibrate convective core overshooting.

\begin{table}[!ht]
\centering
\caption{Atmospheric parameters}
\begin{tabular}{ccccc}
\hline
\hline
																			&	Primary					&	Secondary			\\
\hline
$T_\mathrm{eff}$ (K) 										&	$4930 \pm 30$		&	$6650 \pm 200$	\\
$\log g$ (dex)\tablefootmark{a}						&	$2.91 \pm 0.02$		&	$3.35 \pm 0.02$		\\
$v_\mathrm{t}$ ($\mathrm{km~s^{-1}}$) 	&	$0.98 \pm 0.10 $			&	$3.10 \pm 0.50$				\\
$\mathrm{[Fe/H]}$ (dex)													&	$0.02 \pm 0.05$	&	$-0.05 \pm 0.10$	\\
\hline
\end{tabular}
\tablefoot{a) Effective gravities were kept fixed to the values from \citet{Andersen_1991_06_0}.}
\label{table__Teff}
\end{table}

\subsection{Interferometric observations}

\begin{figure*}[!ht]
\centering
\resizebox{\hsize}{!}{\includegraphics{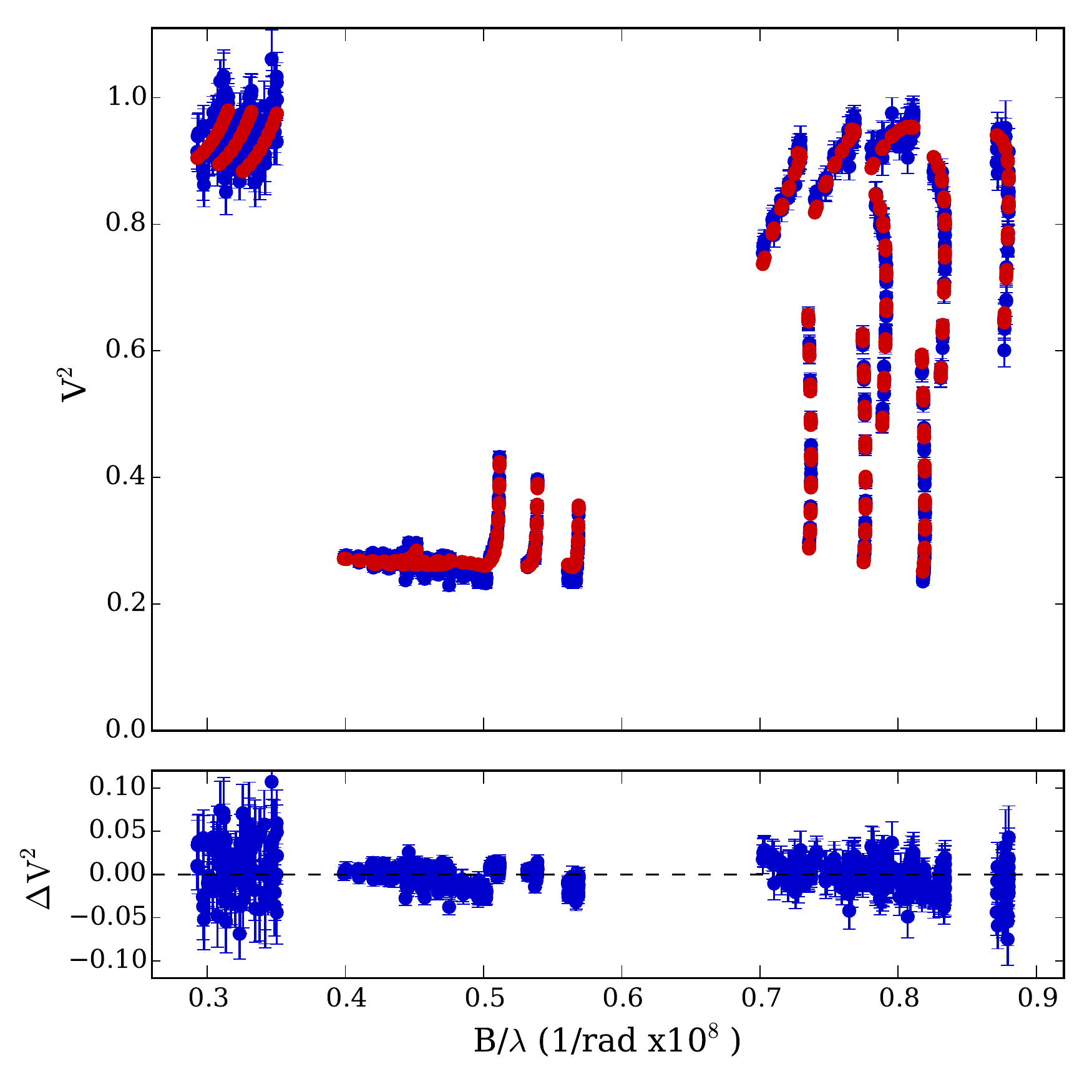}\includegraphics{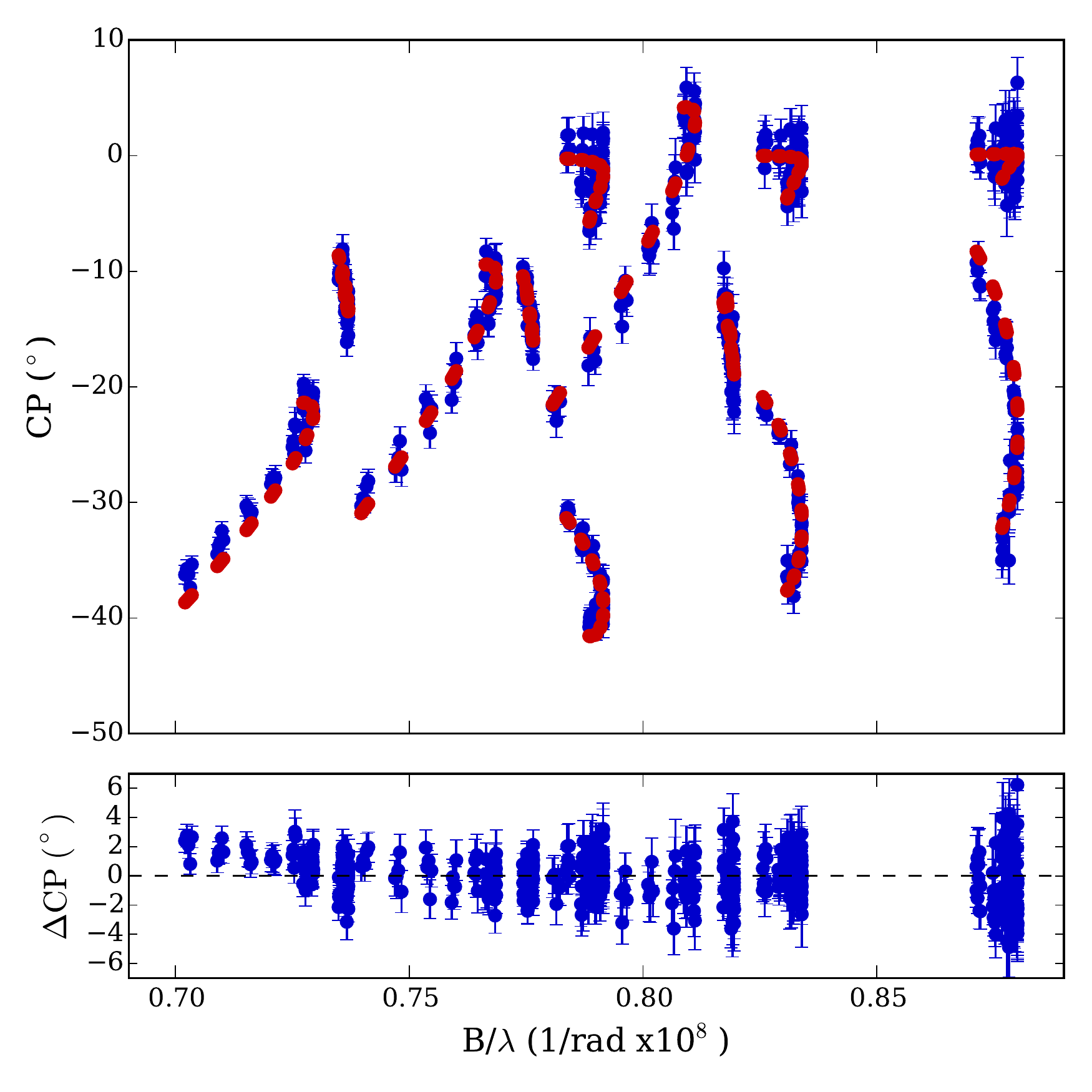}
}
\caption{Squared visibility and closure phase measurements for October 6, 2014. The data are in blue; the red dots are the fitted binary model for this epoch. The residuals are shown at the bottom.}
\label{figure_V2_CP}
\end{figure*}

We used the Very Large Telescope Interferometer \citep[VLTI ;][]{Haguenauer_2010_07_0} with the four-telescope combiner PIONIER \citep{Le-Bouquin_2011_11_0} to measure the squared visibilities and the closure phases. PIONIER combines the light coming from four telescopes in the $H$ band, either in a broadband mode or with a low spectral resolution, where the light is dispersed into several spectral channels. Before December 2014, the instrument allowed a spectral dispersion in three or seven channels; PIONIER was then upgraded (particularly with a grism and a new high-speed infrared detector) and now only offers dispersed fringes in six spectral channels. The recombination from all four telescopes simultaneously provides six visibility and four closure phase measurements across a range of spectral channels.

Our observations were spanned during the ESO period P94 using the 1.8\,m Auxiliary Telescopes with the configuration K0-A1-G1-J3 (baselines of 57, 80, 91, 129, 132, and 140\,m) and  K0-A1-G1-I1 (baselines of 47, 47, 80, 90, 107, and 129\,m), providing six projected baselines ranging from 40 to 140\,m. A log of the observations is listed in Table~\ref{table__astro}. Data were dispersed over three or six spectral channels across the $H$ band ($1.50-1.80\,\mathrm{\mu m}$). To monitor the instrumental and atmospheric contributions, the standard observational procedure, which consists of interleaving the science target by reference stars, was used. The calibrators, listed in Table~\ref{table__astro}, were selected using the \textit{SearchCal}\footnote{Available at http://www.jmmc.fr/searchcal.} software \citep{Bonneau_2006_09_0,Bonneau_2011_11_0} provided by the Jean-Marie Mariotti Center (JMMC).

The data have been reduced with the \textit{pndrs} package described in \citet{Le-Bouquin_2011_11_0}. The main procedure is to compute squared visibilities and triple products for each baseline and spectral channel, and to correct for photon and readout noises. In Fig.~\ref{figure_V2_CP} we present the squared visibilities and closure phases for the first observation, and where we notice that the binary nature of the system is clearly detected. 

In the $H$ band, the primary star (the giant star) has an expected limb-darkened angular diameter $\theta \sim 0.42 \pm 0.02$\,mas \citep{Andersen_1991_06_0}, whereas the secondary has an angular diameter of $0.20 \pm 0.01$\,mas. The primary should be spatially resolved with the larger quadruplet (K0-A1-G1-J3, with three baselines $> 129$\,m), while the secondary is not. We are therefore in the case of a primary resolved star plus a point source component.

For each epoch, we proceeded to a grid search to find the global minimum and the location of the companion. For this we used the interferometric tool \texttt{CANDID}\footnote{Available at \url{https://github.com/amerand/CANDID}} \citep{Gallenne_2015_07_0} to search for the companion using all available observables. \texttt{CANDID} allows a systematic search for point-source companions performing a $N \times N$ grid of fit, whose minimum needed grid resolution is estimated a posteriori. The tool delivers the binary parameters, namely the flux ratio $f$ and the astrometric separation $(\Delta \alpha, \Delta \delta)$, but also the uniform disk angular diameter of the primary $\theta_\mathrm{UD}$ and the detection level of the component. For each epoch, the companion is detected at more than $50\sigma$. The final results are listed in Table~\ref{table__astro}. We estimated the uncertainties from the bootstrapping technique (with replacement) and 10~000 bootstrap samples (also included in the \texttt{CANDID} tool). We then took from the distributions the maximum value between the 16th and 84th percentiles as uncertainty (although the distributions were roughly symmetrical). Finally, the angular diameter was only fitted for the first three observations because we found inconsistent values with the smaller quadruplet. This is likely explained by the fact that the primary is not well resolved with this quadruplet. Its value was therefore fixed to the average value of the first three observations.

The conversion from uniform disk (UD) to limb-darkened (LD) angular diameter was done afterwards by using a linear-law parametrization $I_\lambda (\mu) = 1 - u_\lambda(1 - \mu)$. The LD coefficient $u_\lambda = u_\mathrm{H} = 0.3484$ \citep{Claret_2011_05_0} was chosen taking the stellar parameters $T_\mathrm{eff} = 5000$\,K, $\log g = 3.0$, [Fe/H] $= 0.0$, and $v_\mathrm{t} = 1$\,km~s$^\mathrm{-1}$ (as close as possible to our previous derived values). The conversion is then given by the approximate formula of \citet{Hanbury-Brown_1974_06_0}:
\begin{displaymath}
\theta_\mathrm{LD}(\lambda) = \theta_\mathrm{UD}(\lambda) \sqrt{\frac{1-u_\lambda/3}{1-7u_\lambda/15}}.
\end{displaymath}

We estimated the mean and standard deviation of the first three observations to estimate the LD diameter $\theta_\mathrm{LD} = 0.414 \pm 0.010$\,mas. It is worth mentioning that changing $v_\mathrm{t}$ by $\pm 2$\,km~s$^\mathrm{-1}$ biases the LD diameter by only  0.1\,\% at most, which is well below our precision level.

For the $H$-band flux ratio, we took the mean value and the standard deviation between all observations to obtain $f_\mathrm{H} = 31.8 \pm 0.3$\,\%.

\begin{table*}[!ht]
\centering
\caption{Log of the interferometric observations together with the relative astrometric position of the second component, the flux ratio in the $H$ band and the angular diameter of the primary star.}
\begin{tabular}{ccccccccc}
\hline
\hline
MJD 					& Date & Telescope 		& Spectral & dof & $\Delta x$ &  $\Delta y$ &	$f$ & $\theta_\mathrm{UD}$ \\
						    & 		 & configuration & setup 	&         & (mas)		  &  (mas)          &  (\%)  & (mas) \\
\hline
56937.243  &  2014-Oct-06  &  K0-A1-G1-J3  &  3  channels  & 1349 &  2.419$\pm$0.008  &  0.950$\pm$0.003  &  31.7$\pm$0.5  & 0.391$\pm$0.021  \\
56939.234  &  2014-Oct-08  &  K0-A1-G1-J3  &  3 channels  & 1199 &  2.135$\pm$0.014  &  0.787$\pm$0.004  &  31.5$\pm$0.5  & 0.414$\pm$0.049  \\
56962.235  &  2014-Oct-31  &  K0-A1-G1-J3  &  3 channels  & 1199 &  -2.346$\pm$0.005  &  -1.155$\pm$0.003  &  31.3$\pm$0.6  & 0.405$\pm$0.014  \\
56979.245  &  2014-Nov-17  &  K0-A1-G1-I1  &  3 channels  & 1199 &  -1.801$\pm$0.004  &  -0.615$\pm$0.006  &  31.7$\pm$0.4  & 0.403  \\
56995.141  &  2014-Dec-03  &  K0-A1-G1-I1  &  6 channels  & 149 &  1.547$\pm$0.013  &  0.888$\pm$0.021  &  32.3$\pm$0.5  & 0.403  \\
56996.095  &  2014-Dec-04  &  K0-A1-G1-I1  &  6 channels  & 149 &  1.712$\pm$0.008  &  0.962$\pm$0.008  &  31.4$\pm$0.3  & 0.403  \\
57009.062  &  2014-Dec-17  &  K0-A1-G1-I1  &  6 channels  & 2699 &  2.689$\pm$0.004  &  1.133$\pm$0.007  &  32.2$\pm$0.5  & 0.403  \\
57011.058  &  2014-Dec-19  &  K0-A1-G1-I1  &  6 channels  & 2699 &  2.568$\pm$0.005  &  1.045$\pm$0.008  &  32.2$\pm$0.5  & 0.403  \\
57032.088  &  2015-Jan-09  &  K0-A1-G1-I1  &  6 channels  & 2999 &  -1.382$\pm$0.004  &  -0.830$\pm$0.003  &  32.0$\pm$0.5  & 0.403  \\
57039.099  &  2015-Jan-16  &  K0-A1-G1-I1  &  6 channels  & 2399 &  -2.447$\pm$0.003  &  -1.194$\pm$0.005  &  31.8$\pm$0.5  & 0.403  \\
57048.078  &  2015-Jan-25  &  K0-A1-G1-I1  &  6 channels  & 2398 &  -2.618$\pm$0.002  &  -1.096$\pm$0.003  &  31.8$\pm$0.5  & 0.403  \\
\hline
\end{tabular}
\tablefoot{The listed angular diameter without uncertainty has been fixed to the average value of the three first observations because of a lack of angular resolution with the quadruplet K0-A1-G1-I1. We used the following calibrators: HD~20804: $\theta_\mathrm{UD} = 0.405 \pm 0.029$\,mas, HD~21208: $\theta_\mathrm{UD} = 0.480 \pm 0.034$\,mas, HD~21220: $\theta_\mathrm{UD} = 0.416 \pm 0.030$\,mas, HD~17926: $\theta_\mathrm{UD} = 0.394 \pm 0.038$\,mas, and HD~21882: $\theta_\mathrm{UD} = 0.333 \pm 0.023$\,mas.}
\label{table__astro}
\end{table*}

\section{Orbital solutions, masses, and distance}
\label{section__orbital_solutions_masses_and_distance}

We proceeded to a combined fit of the radial velocities of both components and the astrometry of the relative orbital motion minimizing the $\chi^2$ using a Levenberg-Marquardt method. This provides an unambiguous determination of the masses, distance, and the orbital parameters of the system.

In a Keplerian model, the equations linking the radial velocities $V_1$ and $V_2$ to the relative projected astrometric position ($\Delta \alpha, \Delta \delta$) are \citep{Heintz_1978__0}
\begin{eqnarray*}
\Delta \alpha &=& r [ \sin \Omega \cos(\omega + \nu) + \cos i \cos \Omega \sin(\omega + \nu) ], \\
\Delta \delta &=& r [ \cos \Omega \cos(\omega + \nu) - \cos i \sin \Omega \sin(\omega + \nu) ], \\
r &=& \dfrac{a (1 - e^2)}{1 + e\cos \nu},\\
V_1 &=& \gamma_1 + K_1[\cos(\omega + \nu) + e\cos{\omega}],\\
V_2 &=& \gamma_2 - K_2[\cos(\omega + \nu) + e\cos{\omega}],\\
\end{eqnarray*}
where $a$ is the angular semi-major axis, $e$ the eccentricity, $\Omega$ the longitude of ascending node, $\omega$ the argument of periastron, $i$ the orbital inclination, and $\nu$ the true anomaly estimated from the time, $P_\mathrm{orb}$ the orbital period, and $T_\mathrm{p}$ the time of the spectroscopic conjunction. The parameters $K_1, K_2$ are the radial velocity amplitude of both stars, and $\gamma_1$ and $\gamma_2$ are the systemic velocities for both components. As explained in Sect~\ref{section__observations_and_data_reduction}, we detected an offset of the primary systemic velocity compared to the secondary, we therefore decided to use two different systemic velocities. We then performed a Levenberg-Marquardt fit to our data, whose fitted parameters are $P_\mathrm{orb}, e, \omega, \Omega, i, K_1, K_2, \gamma_1, \gamma_2$ and $a$. The final parameters are listed in Table~\ref{table__results}. To initialize the parameters, we used the values estimated from our spectroscopic analysis and from the published values of \citet{Andersen_1991_06_0}. The value of $T_0$ was kept fixed during the fit.

From these parameters, we can derive the mass of both components and the distance to the system with \citep{Torres_2010_02_0}
\begin{eqnarray*}
M_1 &=& \dfrac{1.036149\times 10^{-7} (K_1 + K_2)^2 K_2 P \sqrt{1 - e^2} }{\sin ^3 i},\\
M_2 &=& \dfrac{1.036149\times 10^{-7} (K_1 + K_2)^2 K_1 P \sqrt{1 - e^2} }{\sin ^3 i},\\
a_\mathrm{AU} &=& \dfrac{9.191966\times 10^{-5} (K_1 + K_2) P \sqrt{1 - e^2} }{\sin i}, \\
d &=& \dfrac{a_\mathrm{AU}}{a}
\end{eqnarray*}
where the masses are expressed in solar units, the distance in parsec, $K_1$ and $K_2$ in km\,s$^{-1}$, $P$ in days, and $a$ in arcsecond. The parameter $a_\mathrm{AU}$ is the linear semi-major axis expressed in astronomical units (the constant value of \citet{Torres_2010_02_0} was converted using the astronomical constants $R_\odot = 695.658 \pm 0.140 \times 10^6$\,m from \citealt{Haberreiter_2008_03_0} and $AU = 149~597~870~700 \pm 3$\,m from \citealt{Pitjeva_2009_04_0}).

\begin{figure*}[!ht]
\centering
\resizebox{\hsize}{!}{\includegraphics{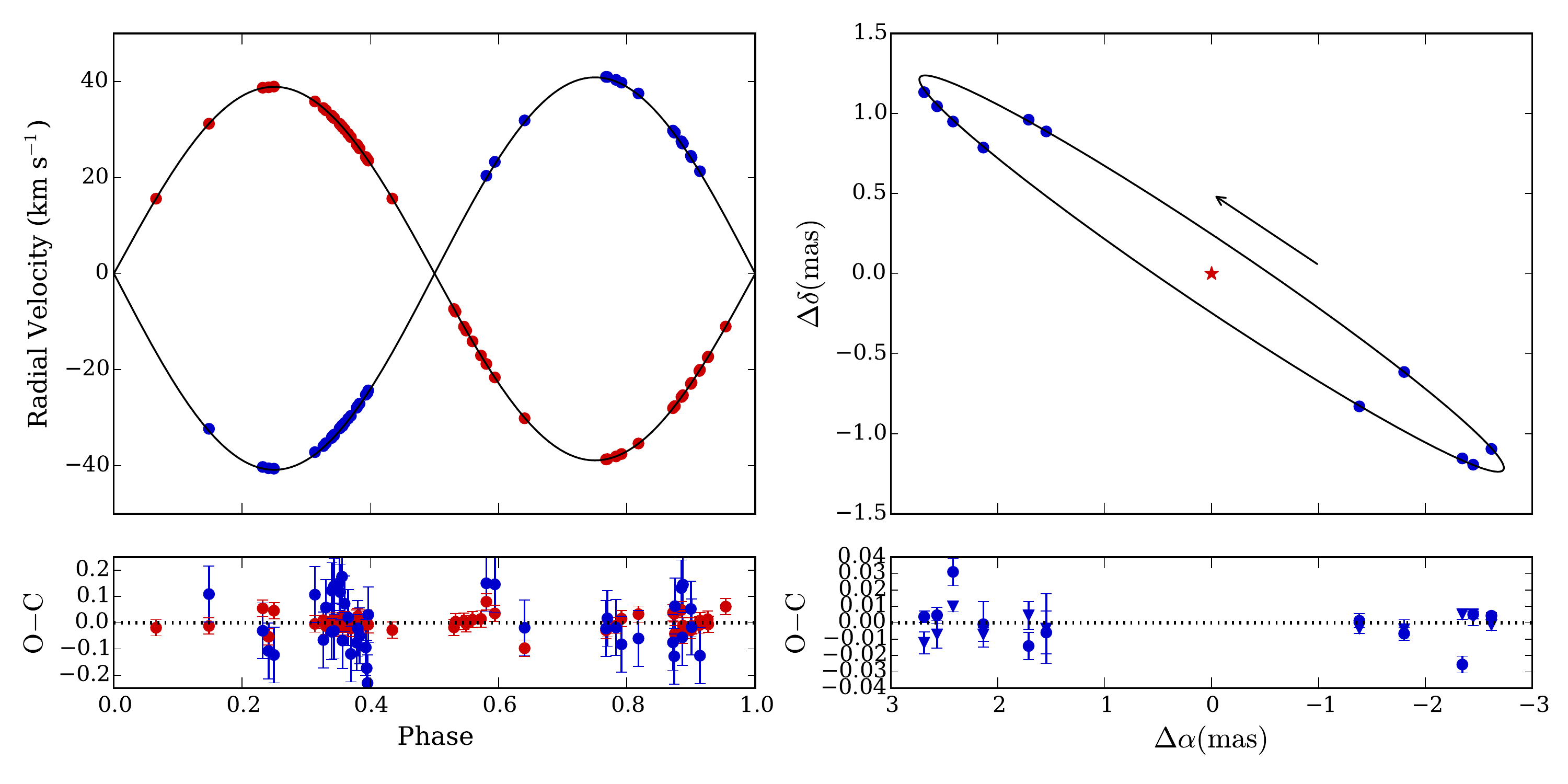}}
\caption{\textit{Left}: Radial velocities of the primary (red) and the secondary (blue). \textit{Right}: Astrometric orbit. The triangles denote the residuals in $\Delta \delta$, while the circles denote the residuals in $\Delta \alpha$.}
\label{figure_orbit}
\end{figure*}

The final derived orbital elements together with the masses and the orbital parallax are listed in Table~\ref{table__results}. The uncertainties were estimated performing 10~000 Monte Carlo simulations. We randomly created synthetic radial velocities and astrometric positions around our best fit solution and for our observing dates. We used normal distributions with standard deviations corresponding to the measurement uncertainties. We then took the median values and used the maximum value between the 16th and 84th percentiles as uncertainty estimates (although the distributions were roughly symmetrical about the median values). The systematic uncertainty from the wavelength calibration \citep[linked to the optical path delay modulation, accurate to 1\,\%,][]{Boffin_2014_04_0} has to be taken into account in the angular size of the orbit (and so the distance); we therefore added in quadrature this 1\,\% uncertainty to the error of the semi-major axis. The final best fit orbit of TZ~For is plotted in Fig.~\ref{figure_orbit}.

\begin{table}[!ht]
\centering
\caption{Best fit orbital elements and parameters. 
}
\begin{tabular}{ccc}
\hline
\hline
Parameter 	& \citet{Andersen_1991_06_0} & This work 	\\
\hline
$P_\mathrm{orb}$ (days)									& $75.6676 \pm0.0010$ & $75.66647 \pm  0.00006$ 			    	\\
$T_\mathrm{p}$ (HJD)									& $2445032.609 \pm 0.002$ &	2452599.29040	 \\
$e$																     & 0.0 &	$0.00002 \pm 0.00003$			   \\
$K_1$ ($\mathrm{km~s^{-1}}$)					  & $38.81 \pm 0.06$ &	$38.90 \pm 0.01$					 	\\
$K_2$ ($\mathrm{km~s^{-1}}$)					  & $40.80 \pm 0.54$ &	$40.87 \pm 0.02$					 	\\
$\gamma_1$	($\mathrm{km~s^{-1}}$)			& $17.42 \pm 0.04$ &	$17.99 \pm 0.03$			 	   			\\
$\gamma_2$	($\mathrm{km~s^{-1}}$)			& $16.30 \pm 0.46$ &	$18.35 \pm 0.11$			 	   			\\
$\omega$	($\degr$)									 & -- &	$269.93 \pm 0.04$				  	\\
$\Omega$	($\degr$)									 & -- &	$65.99 \pm 0.03$			\\
$a$ (mas)\tablefootmark{a}							 & $2.97 \pm 0.18$ &	 $2.993 \pm 0.030$		\\
$a$ (AU)														 & $0.555 \pm 0.004$ &	 $0.5564 \pm 0.0001$		\\
$i$ ($\degr$)													& $85.64 \pm 0.05$ &	$85.68 \pm 0.05$			\\
\hline
$\theta_\mathrm{LD,1}$\tablefootmark{a}  &		$0.418 \pm 0.023$	&		$0.414 \pm	0.010$	\\
$\theta_\mathrm{LD,2}$\tablefootmark{a}  &		$0.199 \pm 0.012$	&		$0.197 \pm	0.008$	\\
$M_1$ ($M_\odot$)							& $2.05 \pm 0.06$ &	$2.057 \pm 0.001$  \\
$M_2$ ($M_\odot$)							& $1.95 \pm 0.03$ &	$1.958 \pm 0.001$   \\
$d$ (pc)											& $185 \pm 10$ &  $185.9 \pm 1.9$    \\
$\pi$ (mas)											& $5.41 \pm 0.29$ &  $5.38 \pm 0.06$    \\
$R_1$ ($R_\odot$)							& $8.32 \pm 0.12$ & $8.28 \pm 0.22$ \\
$R_2$ ($R_\odot$)							& $3.96 \pm 0.09$ & $3.94 \pm 0.17$ \\
$\log L_1/L_\odot$							&	$1.59 \pm 0.04$	&	$1.57 \pm 0.02$ \\
 $\log L_2/L_\odot$							&	$1.36 \pm 0.03$	&	$1.36 \pm 0.03$ \\
\hline& 
\end{tabular}
\tablefoot{\tablefoottext{a}{Estimated from the linear values of \citet{Andersen_1991_06_0}.}}
\label{table__results}
\end{table}

\section{Discussion}
\label{section__discussion}

All derived parameters are in very good agreement with the ones from \citet{Andersen_1991_06_0}, but we reached a better precision, especially for the masses and the distance.

From the measured flux ratio $f_\mathrm{H} = 31.8 \pm 0.3$\,\% and the integrated apparent magnitude of the binary $m_\mathrm{H} = 5.129 \pm 0.024$\,mag \citep{Cutri_2003_03_0}, we estimated an apparent magnitude of the primary and the secondary of $m_\mathrm{H,1} = 5.429 \pm 0.026$\,mag and $m_\mathrm{H,2} = 6.673 \pm 0.027$\,mag, respectively.

The measured angular diameter combined with our derived distance provides a consistent check of the linear radius of the primary, we found $R_1 = 8.28 \pm 0.22\,R_\odot$, which agrees remarkably well (at a $\sim 0.2\sigma$ level) with the value derived by \citet[][$R_1 = 8.32 \pm 0.12\,R_\odot$]{Andersen_1991_06_0}. With the radius ratio $R_1/R_2 = 2.10 \pm 0.07$ estimated by the same authors from photometry, we also derived $R_2 = 3.94\pm 0.17\,R_\odot$ and $\theta_\mathrm{LD} = 0.197 \pm 0.008$\,mas.

The luminosity of both components can also be obtained from
\begin{displaymath}
\dfrac{L_\mathrm{i}}{L_\odot} = \left( \dfrac{R_\mathrm{i}}{R_\odot} \right)^2 \left( \dfrac{T_\mathrm{eff, i}}{T_\odot} \right)^4,
\end{displaymath}
where $\sigma$ is the Stefan–Boltzmann constant, $T_\mathrm{eff}$ the effective temperature, and $i = 1, 2$ for the components. We took $T_\odot = 5777 \pm 10$\,K \citep{Smalley_2005__0}. The effective temperature of the primary is taken from Table~\ref{table__Teff}, while the linear radii and the effective temperature of the secondary is chosen from \citet{Andersen_1991_06_0} because they are more precise. The corresponding luminosities are listed in Table~\ref{table__results}. Our value for the primary is consistent with the previous works.

\begin{figure*}[!ht]
\centering
\resizebox{\hsize}{!}{\includegraphics{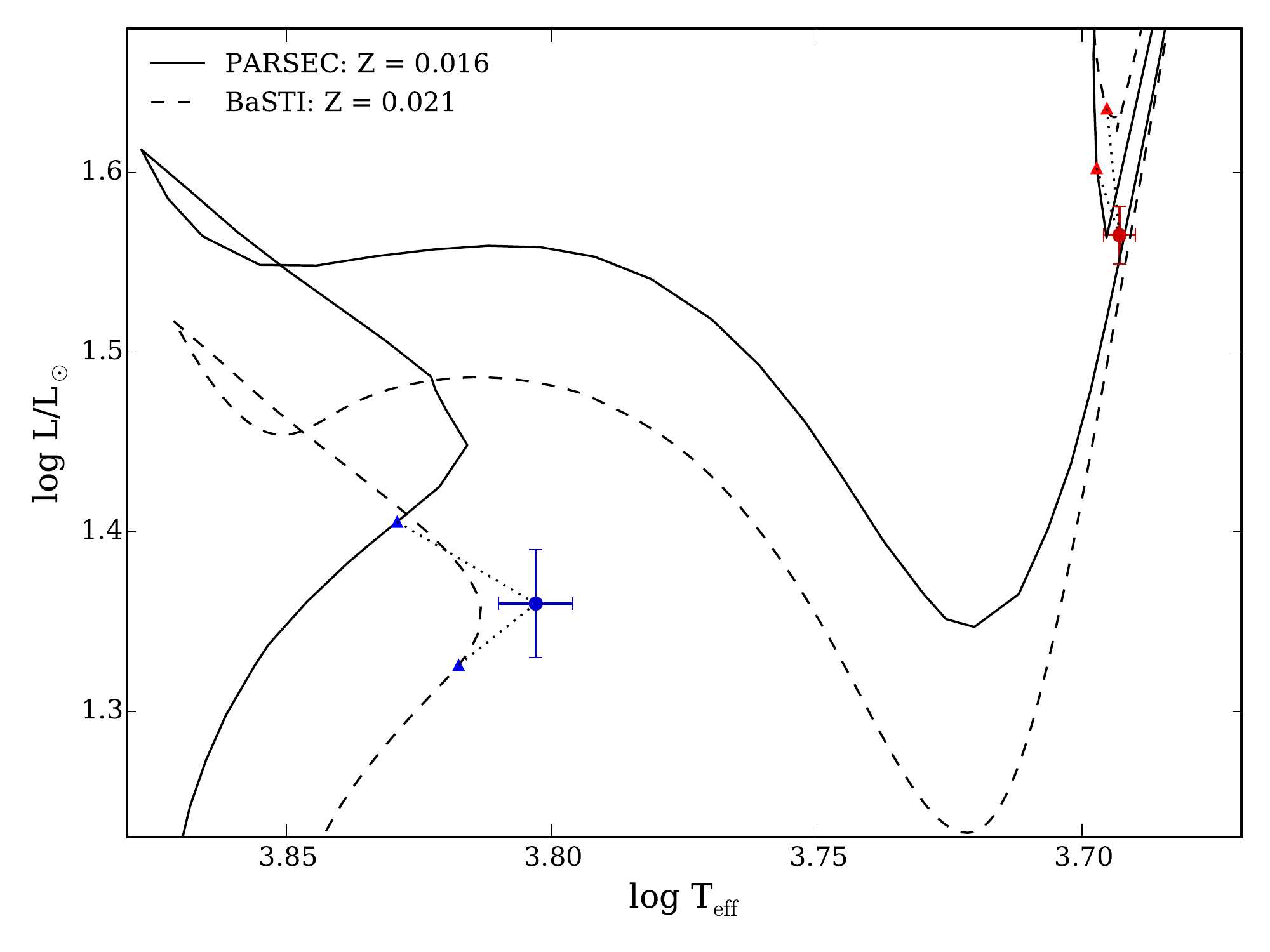}
\includegraphics{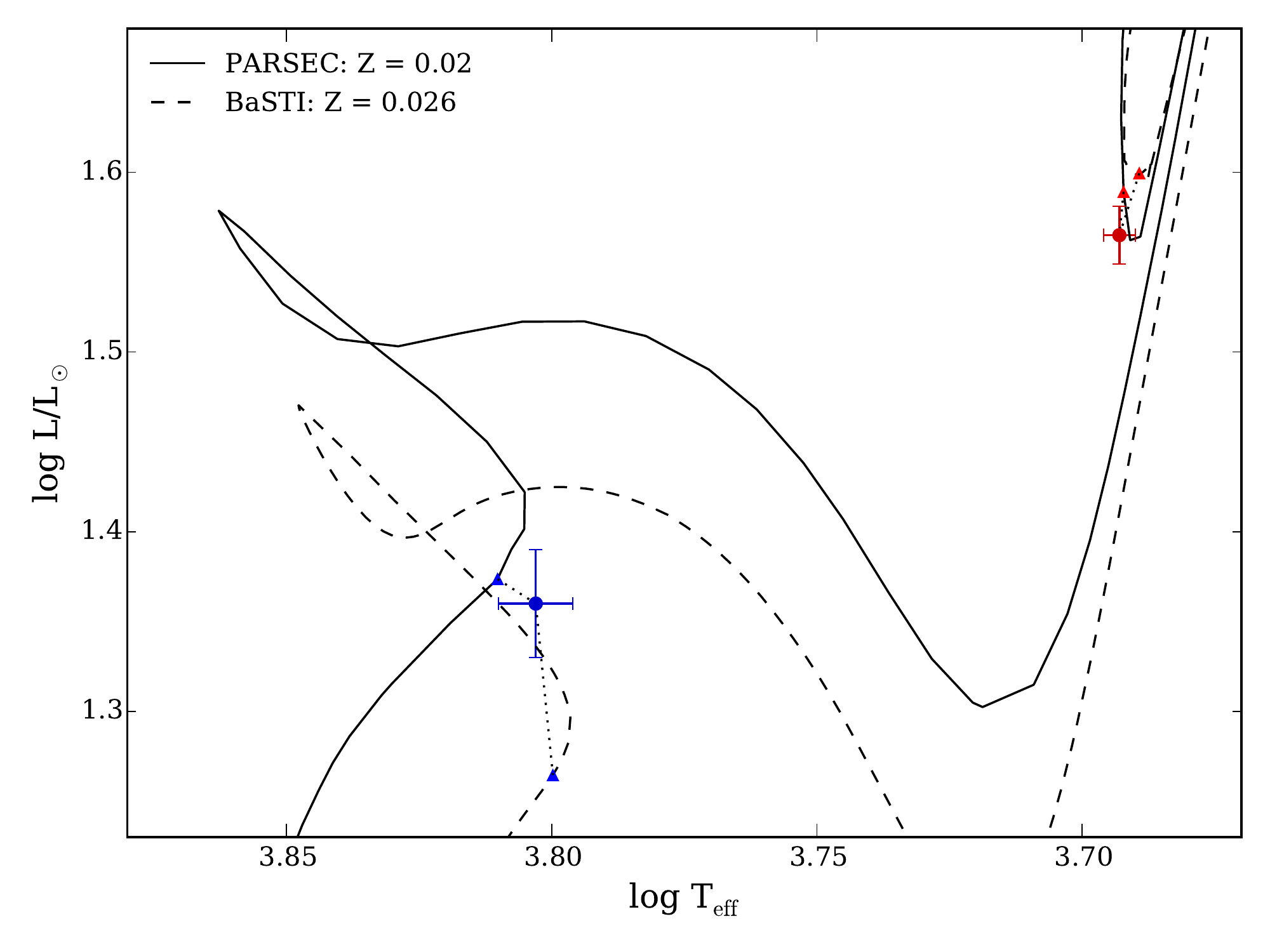}}
\caption{Isochrones from the \texttt{PARSEC} (solid line) and \texttt{BaSTI} (dashed line) stellar evolution models for different metallicities, [Fe/H] = 0.02\,dex (left) and 0.12\,dex (right). The primary is indicated with a red dot, and the secondary in blue. The triangles denote the location of each component according to their corresponding mass.}
\label{figure_isochrones}
\end{figure*}

Our derived masses are also in very good agreement with Andersen et al. results. We improved the relative uncertainties on masses ($< 0.1$\,\%), which are essential for the comparison with stellar evolution models. We checked the evolutionary status of this system using the \texttt{PARSEC} isochrones \citep{Bressan_2012_11_0}. We fitted the age of the system according to our new [Fe/H] estimate, and allowed a single isochrone for both stars. We minimized the $\chi^2$ function including masses, luminosities, effective temperatures, and gravities. We used our derived masses, the effective temperature of the primary listed in Table~\ref{table__Teff}, and the luminosity previously estimated, while the temperature and luminosity of the secondary are taken from \citet{Andersen_1991_06_0} because they are more precise, together with the effective gravities. The value of the primary [Fe/H] = 0.02 were chosen for the models, corresponding to $Z = 0.016$ for $Z_\odot = 0.0152$ (solar metallicity used for the isochrones). We used models that include convective core overshooting with 0.25 pressure scale height above the convective border. A best fit is found for a common age $t = 1.18 \pm 0.03$\,Gyr. We illustrate the solutions in Fig.~\ref{figure_isochrones}. Our inferred age differs by $3.8\sigma$ from the one derived from \citet[$t = 1.75 \pm 0.15$\,Gyr]{Andersen_1991_06_0}, but is more consistent with \citet[$t = 1.3$\,Gyr]{Claret_1995_04_0}, \citet[$t = 1.12-1.33$\,Gyr]{VandenBerg_2006_02_0} and \citet[$t = 1.08-1.41$\,Gyr]{Claret_2011_02_0}. The discrepancy with Andersen's result might be explained by the use of new opacity tables and improved stellar evolution models.

We performed a double check with the \texttt{BaSTI} isochrones \citep{Pietrinferni_2004_09_0}, which also include convective core overshooting with 0.20 pressure scale height and scaled to a solar metallicity $Z_\odot = 0.0198$. This translates to $Z = 0.021$ for our system. We found a best fit for a common age $t = 1.14 \pm 0.04$\,Gyr. This is consistent with the age derived from the \texttt{PARSEC} isochrones. The \texttt{BaSTI} isochrone is plotted in Fig.~\ref{figure_isochrones}. We also note a difference in the HR diagram between the two tracks. This was already pointed out by \citet{Bressan_2012_11_0}. They related the different behaviour of the \texttt{BaSTI} tracks to the use of a different initial solar composition and their calibration to the solar model, resulting in different mixing length parameters (1.91 for \texttt{BaSTI} and 1.74 for \texttt{PARSEC}).

We tried to match the isochrones with different metallicities more closely, and we found that a metalliciy of 0.12\,dex would be better for the \texttt{PARSEC} isochrone with a fitted age $t = 1.24 \pm 0.04$\,Gyr, while there is no improvment with \texttt{BaSTI} and a fitted age $t = 1.23 \pm 0.04$\,Gyr. Isochrones for this metallicity are shown in Fig.~\ref{figure_isochrones}. Even if the metallicity needs to be better constrained, the age seems consistent between various models, we can therefore safely assume an average age of the system to be $t = 1.20 \pm 0.10$\,Gyr. Although the isochrones do not perfectly match the observed position of the components, all models suggest that the primary is in the core helium buring phase. For the secondary, the models suggest that the star is around the ``red hook'', i.e. at the end of its main-sequence phase, and a F7IV-V spectral is therefore more appropriate.

We provided an independent distance measurement which is compatible with the value of \citet{Andersen_1991_06_0} at a 0.5\,\% level. Observations of such systems both by interferometry and photometry will allow the absolute calibration of the SBCR for such stars to be tested, which is of particular importance in determining precise distances ($\sim 1$\,\%) of nearby galaxies from eclipsing binaries \citep{Pietrzynski_2013_03_0,Graczyk_2014_01_0}, and therefore for the extragalactic distance scale.

\begin{figure}[!ht]
\centering
\resizebox{\hsize}{!}{\includegraphics{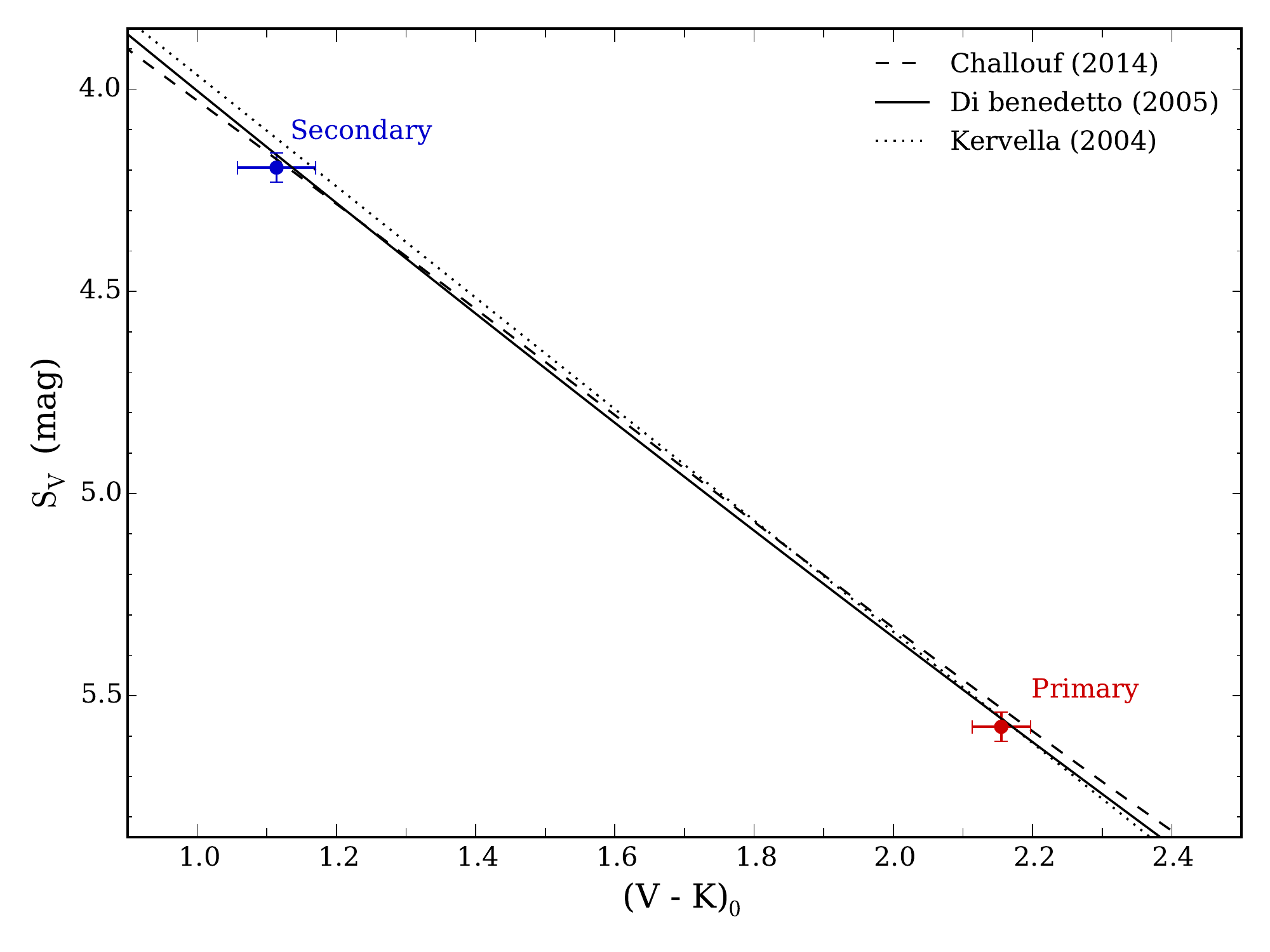}}
\caption{Test of the existing SBC relations using our new precise distance of TZ~For.}
\label{figure_sbcr_test}
\end{figure}

We performed a preliminary test of the existing SBCR used to predict angular diameters and distances. To this end we proceed as follows. The precise semi-major axis of the system given in Table~\ref{table__results} was combined with fractional radii of both components derived by \citet{Andersen_1991_06_0} from light curve analysis. We obtained absolute radii $R_1 = 8.34 \pm 0.11\,R_\odot$ and  $R_2=3.97 \pm 0.08\,R_\odot$. Then we built a model of the system using our Wilson-Devinney code, version 2013 \citep[hereafter WD,][]{Wilson_1971_06_0}, adopting a temperature of the primary component from our atmospheric analysis. The temperature of the secondary was estimated by forcing the $H$-band light ratio calculated with the WD model to be equal to direct interferometric $H$-band flux ratio. The resulting temperature was 6355\,K in good agreement with the value derived by \citet{Andersen_1991_06_0}. We adopted the temperature of the secondary to be $6350 \pm 70$\,K. The slight flattening of the secondary ($\sim 1.2$\,\%) caused by its fast rotation was included in the model. We then used the model to calculate light ratios in Johnson $V$ and $K$ bands. The visible magnitude was taken from \citet{Andersen_1991_06_0} and IR magnitude from 2MASS \citep{Cutri_2003_03_0}; these values were transformed to the Johnson system using equations from \citet{Bessell_1988_09_0} and \citet{Carpenter_2001_05_0}. A possible slight extinction to TZ~For was pointed out previously. We calculated extinction using dust maps from \citet{Schlegel_1998_06_0} with the recalibration from \citet{Schlafly_2011_08_0} to obtain $E(B-V) = 0.015 \pm 0.005$\,mag \citep[details of the method are given in][]{Suchomska_2015_07_0}. Derredened magnitudes and angular diameters of the components were used to calculate the surface brightness $S_\mathrm{V}$ and intrinsic $(V-K)_0$ colours. Figure~\ref{figure_sbcr_test} presents both components of TZ~For against three infrared SBC relations taken from the literature \citet{Challouf_2014_10_0,Di-Benedetto_2005_02_0,Kervella_2004_12_0}. Although all relations are well within observational errors, di Benedetto's relation seems to have the most accurate slope. As di Benedetto's relation is the one used within the Araucaria Project, we confirm here the credibility of our distances to the Magellanic Clouds of our previous works.

\section{Conclusions}
\label{section__conclusion}

We reported the first interferometric observations of the binary system TZ~For using the VLTI/PIONIER combiner. We performed a simultaneous fit of the precise astrometric position obtained from interferometry with new radial velocity measurements. This allowed us to derive the orbital elements, the dynamical masses and the distance to the system with a precision level $< 1.1$\,\%. The values measured for both components were $M_1 = 2.057 \pm 0.001\,M_\odot$ and $M_2 = 1.958 \pm 0.001\,M_\odot$. The comparison with \texttt{PARSEC} and \texttt{BaSTI} isochrones gives an average age of the system of $1.20 \pm 0.10$\,Gyr.

The derived distance is $d = 185.9 \pm 1.9$\,pc, which is a precision of 1.1\,\%, the largest contribution to the distance uncertainty being the accuracy of the wavelength calibration (i.e 1\,\%). This translates to a parallax $\pi = 5.379 \pm 0.055$\,mas. This geometrical and independent estimate of the distance can be used to check the accuracy of other distance indicators, such as the absolute calibration of the surface brightness-colour relation for late-type stars, which is the tool currently used to calibrate the first and most uncertain rungs of the cosmic distance scale using the eclipsing binary method. A preliminary test using our new precise distance of TZ~For shows that the SBCR of \citet{Di-Benedetto_2005_02_0} is the most consistent compared to other published relations. Other systems will be observed in the future using interferometry to check for this calibration with a larger sample.

Finally, independent and precise distance measurements of binary systems combining interferometry and spectroscopy will also provide the best opportunity to check on the future Gaia measurements for possible systematic errors.


\begin{acknowledgements}

The authors would like to thank all the people involved in the VLTI project. We also thank H.~Sana for observing TZ~For as a backup target. A.~G. acknowledges support from FONDECYT grant 3130361. W.~G. and G.~P. gratefully acknowledge financial support for this work from the BASAL Centro de Astrof\'isica y Tecnolog\'ias Afines (CATA) PFB-06/2007. Support from the Polish National Science Centre grants MAESTRO DEC-2012/06/A/ST9/00269 and OPUS DEC-2013/09/B/ST9/01551 is also acknowledged. W.~G. and D.~G. also acknowledge financial support from the Millenium Institute of Astrophysics (MAS) of the Iniciativa Cientifica Milenio del Ministerio de Econom\'ia, Fomento y Turismo de Chile, project IC120009. We also acknowledge support from the ECOS/Conicyt grant C13U01. RIA acknowledges funding from the Swiss National Science Foundation. S.V. acknowledges the support provided by Fondecyt reg.~1130721. The interferometric observations were made with ESO telescopes at Paranal observatory under program ID 094.D-0320. CORALIE observations were made under program CNTAC CN2014A-5. We thank Vincent M\'egevand for technical support at the Swiss telescope. This work made use of the SIMBAD and VIZIER astrophysical database from CDS, Strasbourg, France and the bibliographic information from the NASA Astrophysics Data System. This research has made use of the Jean-Marie Mariotti Center \texttt{SearchCal} and \texttt{ASPRO} services, co-developed by FIZEAU and LAOG/IPAG, and of CDS Astronomical Databases SIMBAD and VIZIER. This work has made use of \texttt{BaSTI} web tools.

\end{acknowledgements}

\bibliographystyle{aa}   
\bibliography{/Users/agallenn/Sciences/Articles/bibliographie}

\end{document}